\documentclass[pra,aps,superscriptaddress,twocolumn]{revtex4}

\usepackage{amsmath}
\usepackage{amsfonts}
\usepackage{bm}
\usepackage{graphicx}

\begin{document}

\title{Ground-state phases of a mixture of spin-1 and spin-2 Bose-Einstein
condensates}

\author{Naoki Irikura}
\affiliation{Department of Engineering Science, University of
Electro-Communications, Tokyo 182-8585, Japan}

\author{Yujiro Eto}
\affiliation{National Institute of Advanced Industrial Science and
Technology (AIST), NMIJ, Tsukuba, Ibaraki 305-8568, Japan}

\author{Takuya Hirano}
\affiliation{Department of Physics, Gakushuin University, Tokyo 171-8588,
Japan}

\author{Hiroki Saito}
\affiliation{Department of Engineering Science, University of
Electro-Communications, Tokyo 182-8585, Japan}

\date{\today}

\begin{abstract}
We investigate the ground-state phases of a mixture of spin-1 and spin-2
Bose-Einstein condensates at zero magnetic field.
In addition to the intra-spin interactions, two spin-dependent interaction
coefficients are introduced to describe the inter-spin interaction.
We systematically explore the wide parameter space, and obtain phase
diagrams containing a rich variety of phases.
For example, there exists a phase in which the spin-1 and spin-2 vectors are
tilted relative to each other breaking the axial symmetry.
\end{abstract}


\maketitle

\section{Introduction}

There is a wide variety of quantum fluids with internal degrees of freedom,
such as superfluid $^3$He~\cite{Vollhardt}, $p$-wave and $d$-wave
superconductors~\cite{Norman}, possible superfluids in neutron
stars~\cite{Hoffberg,Tama}, and spinor Bose-Einstein condensates (BECs) of
atomic gases~\cite{Ho,Ohmi}.
In these systems, the order parameters have spin or angular momentum
degrees of freedom, and their ground-state phases, dynamics, and topological
excitations are richer than those of single-component superfluids.
If two or more quantum fluids with internal degrees of freedom are mixed,
the order-parameter space is greatly extended and the physics is further
enriched.

Spinor BECs of ultracold atoms are suitable systems for realizing such a
mixture of quantum fluids due to their high controllability.
However, in most previous experiments, spinor BECs of spin-1, spin-2, and
spin-3 atoms have been realized only
individually~\cite{Stamper,Stenger,Schmal,Kuwamoto,Naylor}.
The ground state of a spin-1 BEC can be ferromagnetic or
antiferromagnetic, and topological excitations, such as
monopoles~\cite{Ray}, skyrmions~\cite{Leslie}, half-quantum
vortices~\cite{Seo}, and knots~\cite{Hall}, are possible.
A spin-2 BEC is more intriguing because of the presence of the cyclic
phase and non-Abelian vortices~\cite{Koashi,Ciobanu,Kobayashi}.
We expect that a mixture of such spinor BECs will exhibit novel quantum
phases and topological excitations.
A spin-1/spin-1 mixture has been studied theoretically and phase
diagrams and many-body properties have been determined~\cite{Luo, Xu, Shi,
Xu2, Zhang, Xu3, Shi2, Zhang2, Xu4, Xu5, Zhang3}.
The spin dynamics in a mixture of a spin-1 $^{23}$Na BEC and a spin-1
$^{87}$Rb thermal gas have been observed experimentally~\cite{Li}.

Recently, a mixture of spin-1 and spin-2 $^{87}$Rb BECs was realized
experimentally, and the spin dynamics were observed~\cite{Eto}.
Motivated by this experiment, in the present paper we theoretically
investigate the ground-state phase diagrams of the mixture of spin-1 and
spin-2 BECs at zero magnetic field.
The spin-1 and spin-2 BECs have one and two spin-dependent interaction
coefficients, respectively.
In addition to these intra-spin interactions, in a spinor mixture we must
consider the inter-spin interaction, which is described by two spin-dependent
interaction coefficients for the spin-1/spin-2 mixture.
This gives a total of five spin-dependent interaction coefficients.
We therefore study the ground-state phase diagrams by varying these five
interaction coefficients.
Using the Monte Carlo method, we determine the phase diagrams for various
sets of the parameters.
Unlike for the phase diagrams of the individual spin-1 and spin-2 BECs, the
spinor mixture has phases that continuously change with respect to the
interaction coefficients, including phases in which the spin-1 and spin-2
vectors are tilted from each other, breaking the axial symmetry.
According to the interaction coefficients measured in Ref.~\cite{Eto}, the
ground state of the mixture of spin-1 and spin-2 $^{87}$Rb BECs is different
from that of the individual spin-1 and spin-2 BECs.

This paper is organized as follows.
Section~\ref{s:form} presents the problem and reviews the ground states of
spin-1 and spin-2 BECs.
Section~\ref{s:numerical} details the numerical calculations and the various
phase diagrams of the spinor mixture.
Section~\ref{s:conc} provides the conclusions of this study.

\section{Formulation of the Problem}
\label{s:form}

The spin state of spin-1 and spin-2 atoms are denoted by $|f, m\rangle$,
where $f = 1, 2$ and $m = -f, -f + 1, \cdots, f$.
We consider BECs with spin-1 and spin-2 atoms at zero temperature and zero
magnetic field in the mean-field approximation.
The macroscopic wave function for the BEC of spin state $|f, m\rangle$ is
expressed as $\psi^{(f)}_m(\bm{r}) = \sqrt{\rho_f(\bm{r})}
\zeta^{(f)}_m(\bm{r})$, where $\rho_f(\bm{r})$ is the density and
$\zeta^{(f)}_m(\bm{r})$ is the complex spin vector normalized as $\sum_m
|\zeta^{(f)}_m(\bm{r})|^2 = 1$.
The energy $E_f$ of a spin-$f$ BEC with atomic mass $M_f$ confined in a trap
potential $V_f(\bm{r})$ is given by~\cite{Ho,Ohmi,Koashi,Ciobanu}
\begin{eqnarray} \label{E1}
E_1 & = & \int d\bm{r} \sum_{m=-1}^1 \psi^{(1)*}_m(\bm{r}) \left[
-\frac{\hbar^2}{2M_1} \nabla^2 + V_1(\bm{r}) \right] \psi^{(1)}_m(\bm{r})
\nonumber \\
& & + \frac{1}{2} \int d\bm{r} \left[ g_0^{(1)} + g_1^{(1)}
\bm{F}^{(1)}(\bm{r}) \cdot \bm{F}^{(1)}(\bm{r}) \right] \rho_1^2(\bm{r})
\end{eqnarray}
for a spin-1 BEC and
\begin{eqnarray} \label{E2}
E_2 & = & \int d\bm{r} \sum_{m=-2}^2 \psi^{(2)*}_m(\bm{r}) \left[
-\frac{\hbar^2}{2M_2} \nabla^2 + V_2(\bm{r}) \right] \psi^{(2)}_m(\bm{r})
\nonumber \\
& & + \frac{1}{2} \int d\bm{r} \Bigl[ g_0^{(2)} + g_1^{(2)}
\bm{F}^{(2)}(\bm{r}) \cdot \bm{F}^{(2)}(\bm{r})
\nonumber \\
& & + g_2^{(2)} |A_0^{(2)}(\bm{r})|^2 \Bigr] \rho_2^2(\bm{r})
\end{eqnarray}
for a spin-2 BEC, where
\begin{equation}
\bm{F}^{(f)}(\bm{r}) = \sum_{mm'} \zeta_m^{(f)*}(\bm{r}) \bm{S}^{(f)}_{mm'}
\zeta_{m'}^{(f)}(\bm{r})
\end{equation}
is the mean spin vector, with $\bm{S}^{(f)}$ being the vector
of $(2f + 1) \times (2f + 1)$ matrices for spin $f$, and
\begin{equation}
A_0^{(2)} = \frac{1}{\sqrt{5}} \left( 2 \zeta_2^{(2)} \zeta_{-2}^{(2)} - 2
\zeta_1^{(2)} \zeta_{-1}^{(2)} + \zeta_0^{(2)2} \right)
\end{equation}
is the spin-singlet scalar for spin 2.
The interaction coefficients in Eqs.~(\ref{E1}) and (\ref{E2}) have the
forms
\begin{eqnarray}
g_0^{(1)} & = & \frac{4\pi\hbar^2}{M_1} \frac{a_0^{(1)} + 2 a_2^{(1)}}{3},
\\
g_1^{(1)} & = & \frac{4\pi\hbar^2}{M_1} \frac{a_2^{(1)} - a_0^{(1)}}{3},
\\
g_0^{(2)} & = & \frac{4\pi\hbar^2}{M_2}
\frac{4 a_2^{(2)} + 3 a_4^{(2)}}{7}, \\
g_1^{(2)} & = & \frac{4\pi\hbar^2}{M_2} \frac{a_4^{(2)} - a_2^{(2)}}{7},
\\
g_2^{(2)} & = & \frac{4\pi\hbar^2}{M_2}
\frac{7 a_0^{(2)} - 10 a_2^{(2)} + 3 a_4^{(2)}}{7},
\end{eqnarray}
where $a_{\cal F}^{(f)}$ is the $s$-wave scattering length between
spin-$f$ atoms with colliding channel of total spin ${\cal F}$.
We denote the spin vectors as $\bm{\zeta}^{(1)} = (\zeta_1^{(1)},
\zeta_0^{(1)}, \zeta_{-1}^{(1)})$ and $\bm{\zeta}^{(2)} = (\zeta_2^{(2)},
\zeta_1^{(2)}, \zeta_0^{(2)}, \zeta_{-1}^{(2)}, \zeta_{-2}^{(2)})$.

\begin{figure}[tb]
\includegraphics[width=8.5cm]{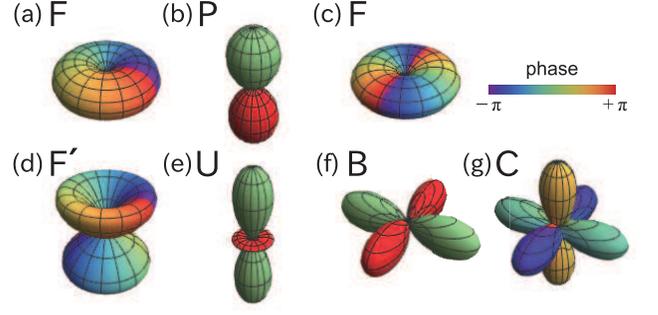}
\caption{
(color online) Spherical harmonic representations $S(\theta, \phi)$ of the
spin states.
(a) spin-1 ferromagnetic state $\bm{\zeta}^{(1)}_F$.
(b) spin-1 polar state $\bm{\zeta}^{(1)}_P$.
(c) spin-2 ferromagnetic state $\bm{\zeta}^{(2)}_F$.
(d) spin-2 state $\bm{\zeta}^{(2)}_{F'}$.
(e) spin-2 uniaxial nematic state $\bm{\zeta}^{(2)}_{\rm UN}$.
(f) spin-2 biaxial nematic state $\bm{\zeta}^{(2)}_{\rm BN}$.
(g) spin-2 cyclic state $\bm{\zeta}^{(2)}_C$.
The labels F, P, F$'$, U, B, and C shown for each representation are used to
identify the spin state in the phase diagram.
}
\label{f:spin12}
\end{figure}

Before considering the mixture of spinor BECs, we summarize the
ground-state phases for individual spin-1 and spin-2 BECs in a uniform
system.
The ground state of a spin-1 BEC depends on the sign of $g_1^{(1)}$.
When $g_1^{(1)} < 0$, the ground state is the fully-polarized
ferromagnetic state
\begin{equation} \label{f1}
\bm{\zeta}^{(1)}_F \equiv e^{i\chi} \hat R (1, 0, 0),
\end{equation}
where $\chi$ is an arbitrary phase and $\hat R$ is an arbitrary SO(3)
rotation in the spin space.
When $g_1^{(1)} > 0$, the ground state is the polar state
\begin{equation} \label{p1}
\bm{\zeta}^{(1)}_P \equiv e^{i\chi} \hat R (0, 1, 0).
\end{equation}
The spin-2 BEC has more variety of ground states.
When $g_1^{(2)} < 0$ and $g_2^{(2)} > 20 g_1^{(2)}$, the ground state is
the ferromagnetic state
\begin{equation} \label{f2}
\bm{\zeta}^{(2)}_F \equiv e^{i\chi} \hat R (1, 0, 0, 0, 0).
\end{equation}
When $g_2^{(2)} < 0$ and $g_2^{(2)} < 20 g_1^{(2)}$, the ground state has
continuous degeneracy: a linear combination of the uniaxial nematic state
\begin{equation} \label{u2}
\bm{\zeta}^{(2)}_{\rm UN} \equiv e^{i\chi} \hat R (0, 0, 1, 0, 0)
\end{equation}
and the biaxial nematic state
\begin{equation} \label{b2}
\bm{\zeta}^{(2)}_{\rm BN} \equiv e^{i\chi} \hat R (1, 0, 0, 0, 1) / \sqrt{2}
\end{equation}
is the ground state.
When $g_1^{(2)} > 0$ and $g_2^{(2)} > 0$, the ground state is the cyclic
state
\begin{equation} \label{c2}
\bm{\zeta}^{(2)}_C \equiv e^{i\chi} \hat R (1 / 2, 0, i / \sqrt{2}, 0, 1 /
2).
\end{equation}
For later use, we define the state
\begin{equation} \label{f2p}
\bm{\zeta}^{(2)}_{F'} \equiv e^{i\chi} \hat R (0, 1, 0, 0, 0),
\end{equation}
which is not the ground state
but a stationary state of the Gross-Pitaevskii equation.
The spherical harmonic representation of the spin state is convenient for
visualizing the symmetry of the system~\cite{Kawaguchi},
\begin{equation}
S(\theta, \phi) = \sum_{m = -f}^f \zeta_m^{(f)} Y_f^m(\theta, \phi),
\end{equation}
where $Y_f^m$ is the spherical harmonics.
The spherical harmonic representations of the above spin states are shown
in Fig.~\ref{f:spin12}.

We consider a mixture of spin-1 and spin-2 BECs.
The interaction energy between the spin-1 and spin-2 BECs is obtained to be
(see Appendix for derivation)
\begin{eqnarray} \label{E12}
E_{12} & = & \int d\bm{r} \Bigl[ g_0^{(12)}
+ g_1^{(12)} \bm{F}^{(1)}(\bm{r}) \cdot \bm{F}^{(2)}(\bm{r})
\nonumber \\
& & + g_2^{(12)} P_1^{(12)}(\bm{r}) \Bigr] \rho_1(\bm{r}) \rho_2(\bm{r}),
\end{eqnarray}
where $P_1^{(12)}$ is defined in Eq.~(\ref{P1}).
The interaction coefficients in Eq.~(\ref{E12}) are given by
\begin{subequations} \label{g12}
\begin{eqnarray}
g_0^{(12)} & = & \frac{2\pi\hbar^2}{M_{12}}
\frac{2 a_2^{(12)} + a_3^{(12)}}{3}, \\
g_1^{(12)} & = & \frac{2\pi\hbar^2}{M_{12}}
\frac{a_3^{(12)} - a_2^{(12)}}{3}, \\
g_2^{(12)} & = & \frac{2\pi\hbar^2}{M_{12}}
\frac{3 a_1^{(12)} - 5 a_2^{(12)} + 2 a_3^{(12)}}{3},
\end{eqnarray}
\end{subequations}
where $M_{12} = (M_1^{-1} + M_2^{-1})^{-1}$ is the reduced mass and
$a_{\cal F}^{(12)}$ is the $s$-wave scattering length between spin-1 and
spin-2 atoms with colliding channel of total spin ${\cal F}$.

In the following analysis, we assume that the spin healing lengths are much
larger than the size of the atomic cloud and we neglect the spatial
variation of the spin states $\bm{\zeta}^{(f)}$.
The kinetic and potential energy terms in $E_1$ and $E_2$ in
Eqs.~(\ref{E1}) and (\ref{E2}) then become independent of the spin states
$\bm{\zeta}^{(f)}$.
The spin-dependent part of the total energy $E = E_1 + E_2 + E_{12}$ thus
reduces to
\begin{eqnarray} \label{Espin}
E_{\rm spin} & = & \frac{1}{2} \left(
c_1^{(1)} \bm{F}^{(1)} \cdot \bm{F}^{(1)} + c_1^{(2)} \bm{F}^{(2)} \cdot
\bm{F}^{(2)} + c_2^{(2)} |A_0^{(2)}|^2 \right)
\nonumber \\
& & + c_1^{(12)} \bm{F}^{(1)} \cdot \bm{F}^{(2)} + c_2^{(12)} P_1^{(12)},
\end{eqnarray}
where
\begin{eqnarray}
c_n^{(f)} & = & g_n^{(f)} \int \rho_f^2(\bm{r}) d\bm{r},
\nonumber \\  
c_n^{(12)} & = & g_n^{(12)} \int \rho_1(\bm{r}) \rho_2(\bm{r}) d\bm{r}
\label{cdef}
\end{eqnarray}
with $n = 1, 2$.
In the rest of this paper, we normalize the interaction coefficients
$c_1^{(1)}$, $c_1^{(2)}$, $c_2^{(2)}$, $c_1^{(12)}$, and $c_2^{(12)}$ by
$4\pi \hbar^2 a_B \int \rho_1^2 d\bm{r} / M_1$, where $a_B$ is the Bohr
radius, and therefore these interaction coefficients are dimensionless.

Our purpose is to find the spin states $\bm{\zeta}^{(1)}$ and
$\bm{\zeta}^{(2)}$ that minimize the energy $E_{\rm spin}$.
We numerically obtain the ground state as follows.
First we set complex random numbers to $\zeta_m^{(f)}$ and minimize the
energy in a stochastic manner, that is, we try a small random change to the
spin state $\zeta_m^{(f)} + \delta\zeta_m^{(f)}$ and adopt the change if
the energy is lowered.
After sufficiently many steps in this random walk in the spin space, we
obtain a metastable state or the ground state.
Repeating this procedure many times with different initial random states, we
can exclude metastable states and determine the true ground state.

\section{Ground states of a spin-1/spin-2 mixture}
\label{s:numerical}

\begin{figure}[tb]
\includegraphics[width=8.5cm]{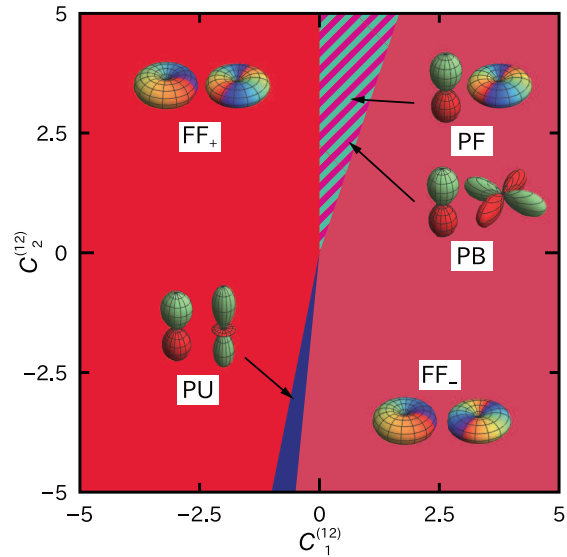}
\caption{
(color online) Ground-state phase diagram with respect to the inter-spin
interactions $c_1^{(12)}$ and $c_2^{(12)}$ without the intra-spin
interactions, $c_1^{(1)} = c_1^{(2)} = c_2^{(2)} = 0$.
The spherical harmonic representations of the spin states are also shown,
where the left- and right-hand figures indicate the spin-1 and spin-2
states, respectively.
The letter-pairs specify the spin-1 and spin-2 states, which are defined in
Fig.~\ref{f:spin12}, and the subscript $\pm$ indicates the sign of
$\bm{F}^{(1)} \cdot \bm{F}^{(2)}$.
In the striped region, the linear combination of the polar-ferromagnetic and
polar-biaxial nematic states are continuously degenerate.
}
\label{f:noint}
\end{figure}
\begin{figure}[tb]
\includegraphics[width=7.6cm]{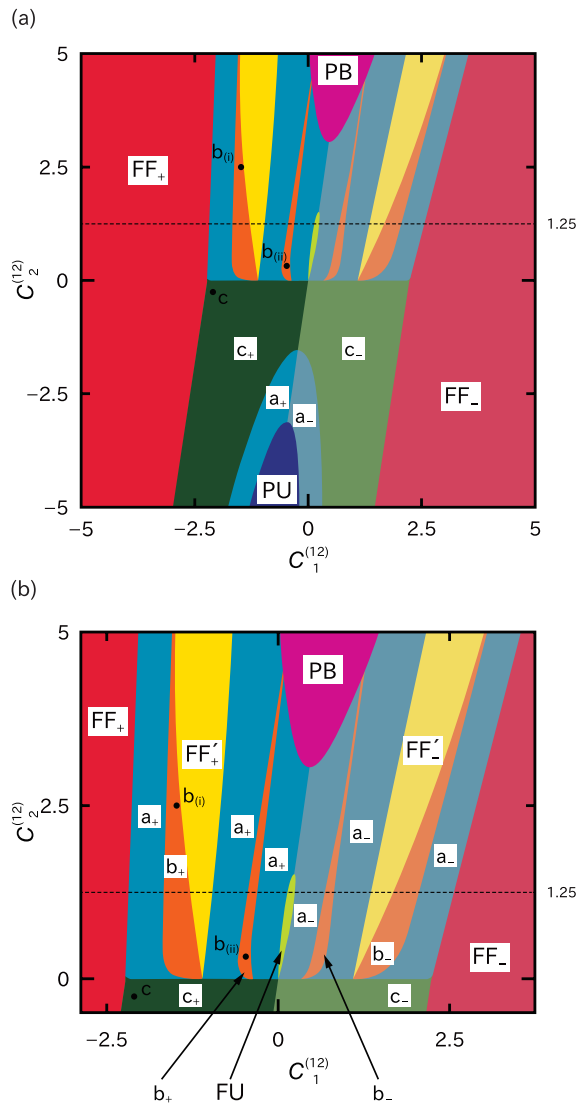}
\caption{
  (color online) Ground-state phase diagram for $c_1^{(1)} = -0.46$,
  $c_1^{(2)} = 1.1$, and $c_2^{(2)} = -0.05$.
  The ground state for $c_1^{(12)} = c_2^{(12)} = 0$ is the ferromagnetic
  state for spin 1 and the nematic state for spin 2.
  The region of many phases in (a) is magnified in (b).
  The upper-case letter-pairs indicate the spin-1 and spin-2 states as
  defined in Fig.~\ref{f:spin12} and the lower-case letters indicate the
  intermediate states as defined in Table~\ref{t:abc}.
  The subscripts $\pm$ denote the sign of $\bm{F}^{(1)} \cdot
  \bm{F}^{(2)}$.
  The physical quantities along the dotted line are shown in
  Fig.~\ref{f:ferro_nem2}(a).
  The spin states at the black dots are shown in
  Fig.~\ref{f:ferro_nem2}(b).
 }
\label{f:ferro_nem}
\end{figure}
To see the effect of the interaction between the spin-1 and spin-2 BECs, we
first consider the case without the intra-spin interactions,
$c_1^{(1)} = c_1^{(2)} = c_2^{(2)} = 0$.
The spin-dependent energy then reduces to $E_{\rm spin} = c_1^{(12)}
\bm{F}^{(1)} \cdot \bm{F}^{(2)} + c_2^{(12)} P_1^{(12)}$.
Figure~\ref{f:noint} shows the ground-state phase diagram with respect to
$c_1^{(12)}$ and $c_2^{(12)}$.
When $c_1^{(12)}$ is sufficiently large and negative, the state in which the
spin vectors $\bm{F}^{(1)}$ and $\bm{F}^{(2)}$ are fully-polarized in the
same direction is energetically favored, and the ground state is
$\bm{\zeta}^{(1)} = \bm{\zeta}^{(1)}_F$ and $\bm{\zeta}^{(2)} =
\bm{\zeta}^{(2)}_F$.
We abbreviate this state as ``FF$_+$'', in which the first and second
letters indicate the spin-1 and spin-2 states, respectively, and the
subscript $+$ denotes that the two spin vectors are in the same direction.
The capital letters indicate the spin states shown in Fig.~\ref{f:spin12}.
The energy of the FF$_+$ state is $E_{\rm spin} = 2 c_1^{(12)}$.
In a similar manner, when $c_1^{(12)}$ is large and positive, the ground
state is the ferromagnetic state with $\bm{F}^{(1)}$ and $\bm{F}^{(2)}$
being in opposite directions, whose energy is $E_{\rm spin} = -2 c_1^{(12)}
+ 3 c_2^{(12)} / 5$.
This phase is denoted as ``FF$_-$'', where the subscript $-$ represents that
the two spin vectors are in the opposite directions.
In general, we define the subscripts $\pm$ to indicate the sign of
$\bm{F}^{(1)} \cdot \bm{F}^{(2)}$.
As shown in Fig.~\ref{f:noint}, there are two regions between these
ferromagnetic phases.
When $c_2^{(12)} < 0$ and $c_2^{(12)} / 5 < c_1^{(12)} < c_2^{(12)} /
10$, the ground state is $\bm{\zeta}^{(1)} = \bm{\zeta}^{(1)}_P$ and
$\bm{\zeta}^{(2)} = \bm{\zeta}^{(2)}_{\rm UN}$ with an energy $E_{\rm spin}
= 2 c_2^{(12)} / 5$, which is denoted as ``PU''.
When $c_2^{(12)} > 0$ and $0 < c_1^{(12)} < 3 c_2^{(12)} / 10$, the ground
state is continuously degenerate: the linear combination of the ``PF''
(polar-ferromagnetic) and ``PB'' (polar-biaxial nematic) states is the ground
state with an energy $E_{\rm spin} = 3 c_2^{(12)} / 10$.

\begin{table}
  \begin{tabular}{l|lllll}
    & $|\bm{F}^{(1)}|$ & $|\bm{F}^{(2)}|$ & $A_0^{(2)}$ & $\bm{F}^{(1)}
    \times \bm{F}^{(2)}$ & isotropy group \\ \hline
    a & nonzero & nonzero & nonzero & 0 & $\mathbb{Z}_2$ \\
    b & nonzero & nonzero & nonzero & nonzero & E \\
    c & 1 & nonzero & nonzero & 0 & $\mathbb{Z}_4$ \\
    d & 1 & nonzero & 0 & 0 & $\mathbb{Z}_3$ \\
    e & 0 & 0 & nonzero & 0 & $\mathbb{Z}_2 \times \mathbb{Z}_2$
  \end{tabular}
  \caption{
    Classification of the intermediate states that change continuously in
    the phase diagram.
    ``nonzero'' indicates that the value depends on $c_1^{(12)}$ and
    $c_2^{(12)}$.
    E indicates the trivial group.
    The subscript $+$ or $-$ is added to a-d to indicate the sign of
    $\bm{F}^{(1)} \cdot \bm{F}^{(2)}$.
  }
  \label{t:abc}
\end{table}
Next we consider the cases of nonzero intra-spin interaction coefficients
$c_1^{(1)}$, $c_1^{(2)}$, and $c_2^{(2)}$.
Figure~\ref{f:ferro_nem} shows the ground-state phase diagram for $c_1^{(1)}
= -0.46$, $c_1^{(2)} = 1.1$, and $c_2^{(2)} = -0.05$, which correspond to
the interaction coefficients of $^{87}$Rb for $\rho_1 = \rho_2$ in
Eq.~(\ref{cdef}).
There is a remarkable number of phases with complicated structures.
If the inter-spin interaction is absent, i.e., at the origin of the phase
diagram, the ground state for spin 1 is the ferromagnetic state and that for
spin 2 is the nematic state.
Comparing Fig.~\ref{f:ferro_nem} with Fig.~\ref{f:noint}, we find that the
four phases in Fig.~\ref{f:noint}, FF$_+$, FF$_-$, PU, and PB, also appear
in Fig.~\ref{f:ferro_nem}, where the continuous degeneracy in
Fig.~\ref{f:noint} is removed and the PF state disappears in
Fig.~\ref{f:ferro_nem}.
There are many intermediate states, labeled by lower-case letters classified
in Table~\ref{t:abc}.
In the regions of these intermediate states, either or both of the spin-1
and spin-2 states continuously change with respect to $c_1^{(12)}$ and
$c_2^{(12)}$.

\begin{figure}[tb]
\includegraphics[width=8.5cm]{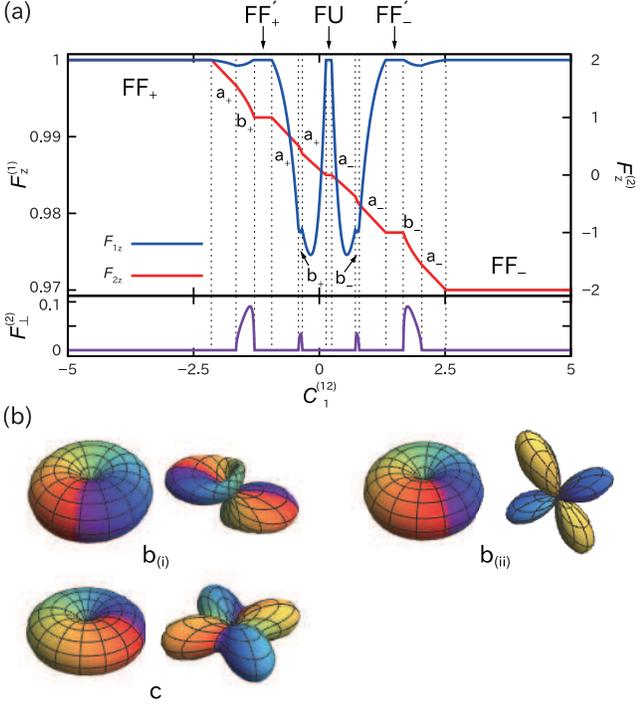}
\caption{
  (color online) (a) Dependence of $F_z^{(1)}$, $F_z^{(2)}$, and
  $F_\perp^{(2)}$ on $c_1^{(12)}$ along the dotted line in
  Fig.~\ref{f:ferro_nem}, where $F_\perp^{(f)}  = [(F_x^{(f)})^2 +
  (F_y^{(f)})^2]^{1/2}$.
  Here, the spin-1 and spin-2 states are rotated so that $\bm{F}^{(1)}$ is
  in the $z$ direction, and hence $F_\perp^{(1)}$ is always zero.
  (b) Spherical-harmonic representations of the spin states marked by the
  black dots in Fig.~\ref{f:ferro_nem}, where the left- and right-hand
  figures are the spin-1 and spin-2 states, respectively.
 }
\label{f:ferro_nem2}
\end{figure}
We now consider the phases along the dotted line in Fig.~\ref{f:ferro_nem}.
When $c_1^{(12)}$ is large and negative, the ground state is the FF$_+$
state.
When $c_1^{(12)}$ crosses the phase boundary between FF$_+$ and a$_+$, the
lengths of the spin-1 and spin-2 vectors begin to decrease, as shown in
Fig.~\ref{f:ferro_nem2}(a).
In this a$_+$ phase, the spin vectors $\bm{F}^{(1)}$ and $\bm{F}^{(2)}$
remain in the same direction.
In contrast, in the b$_+$ phase, the directions of the spin vectors
$\bm{F}^{(1)}$ and $\bm{F}^{(2)}$ become different.
This can be regarded as axisymmetry breaking of the magnetization, that is,
if we fix the vector $\bm{F}^{(1)}$ to the $z$ direction, the vector
$\bm{F}^{(2)}$ has a component $F_\perp^{(2)}$ perpendicular to the $z$
axis.
Examples of such axisymmetry breaking states are shown in
Fig.~\ref{f:ferro_nem2}(b).
Axisymmetry breaking has been found in a spin-1/spin-1 mixture in the
presence of an external magnetic field~\cite{Xu3}.
In the FF$'_+$ phase, the directions of the spin vectors $\bm{F}^{(1)}$ and
$\bm{F}^{(2)}$ become the same again.
In this phase, the spin-1 state returns to $\bm{\zeta}^{(1)} =
\bm{\zeta}^{(1)}_F$ and the spin-2 state is $\bm{\zeta}^{(2)} =
\bm{\zeta}^{(2)}_{F'}$, which does not depend on $c_1^{(12)}$ and
$c_2^{(12)}$ within the phase, as seen from the plateau in
Fig.~\ref{f:ferro_nem2}(a).
In the a$_+$, b$_+$, and a$_+$ phases, the spin states continuously change
again; the spin vectors $\bm{F}^{(1)}$ and $\bm{F}^{(2)}$ are in the same
direction in the a$_+$ phase, while they take different directions in the
b$_+$ phase.
The FU state is connected to the origin of the phase diagram.
The phases on the right-hand side of the phase diagram, a$_-$, b$_-$,
$\cdots$ are similar to the corresponding phases a$_+$, b$_+$, $\cdots$
where the spin vector $\bm{F}^{(1)}$ or $\bm{F}^{(2)}$ is flipped, i.e., the
time-reversal transformation is applied to the spin-1 or spin-2 state.
For example, in the FF$'_-$ phase, when the spin-1 state is
$\bm{\zeta}^{(1)} = (1, 0, 0)$, the spin-2 state is $\bm{\zeta}^{(2)} = (0,
0, 0, 1, 0)$, which is the time-reversal state of $\bm{\zeta}^{(2)} = (0, 1,
0, 0, 0)$ in the FF$'_+$ phase.
For $c_2^{(12)} < 0$, the phase structures are simpler.
In the c$_\pm$ phases, the spin-1 state is fixed to the ferromagnetic
state, while the spin-2 state continuously changes with $\bm{F}^{(1)}$ and
$\bm{F}^{(2)}$ being kept in the same direction.
A typical $c$ state is shown in Fig.~\ref{f:ferro_nem2}(b).

In the experiment in Ref.~\cite{Eto}, the values of the inter-spin
scattering lengths of $^{87}$Rb were measured, which correspond to
$c_1^{(12)} \simeq 0.83$ and $c_2^{(12)} \simeq 4.8$ in the present case, if
$\rho_1 = \rho_2$ in Eq.~(\ref{cdef}), i.e., an almost 1:1 mixture of spin-1
and spin-2 atoms.
In the phase diagram in Fig.~\ref{f:ferro_nem}, these values correspond to
the PB state, namely, the polar state for spin 1 and the biaxial nematic
state for spin 2.
The ground state phase of the spin-1 $^{87}$Rb BEC alone is the
ferromagnetic state and that for spin-2 is the biaxial or uniaxial nematic
state.
Thus, the ground state of the 1:1 mixture of spin-1 and spin-2 $^{87}$Rb
BECs is different from those of the individual BECs due to the inter-spin
interaction.

\begin{figure}[tb]
\includegraphics[width=8cm]{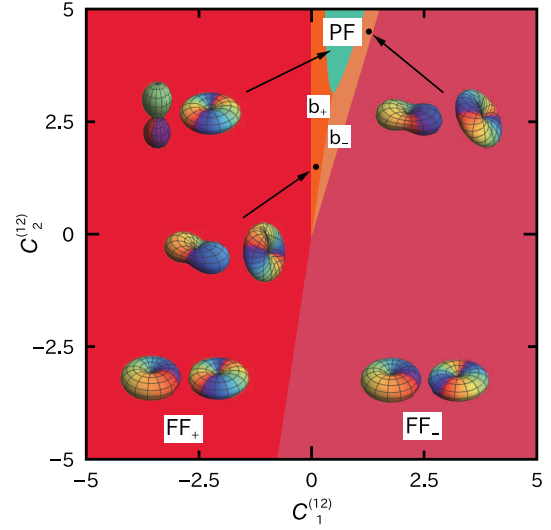}
\caption{
  (color online) Ground-state phase diagram for $c_1^{(1)} = -0.46$,
  $c_1^{(2)} = -1.1$, and $c_2^{(2)} = 1.5$.
  The ground state for $c_1^{(12)} = c_2^{(12)} = 0$ is the ferromagnetic
  state for both spin 1 and spin 2.
  The spherical harmonic representations of the spin states are also shown,
  where the left- and right-hand figures represent the spin-1 and spin-2
  states, respectively.
 }
\label{f:ferro_ferro}
\end{figure}
Figure~\ref{f:ferro_ferro} shows the ground-state phase diagram for
$c_1^{(1)} = -0.46$, $c_1^{(2)} = -1.1$, and $c_2^{(2)} = 1.5$.
If the inter-spin interaction is absent, the ground state is the
ferromagnetic state both for spin 1 and spin 2 for these parameters.
The phase diagram is much simpler than Fig.~\ref{f:ferro_nem}.
Comparing Fig.~\ref{f:ferro_ferro} with Fig.~\ref{f:noint}, we find that the
PB and PU states disappear in Fig.~\ref{f:ferro_ferro}.
Between the PF and FF$_\pm$ phases, there exists the region of the b state,
in which the axisymmetry is broken.
For the present parameters, the spin 2 state is almost the ferromagnetic
state in the b phase.
The angle between the two spin vectors changes from 0 to $\pi$ across the
region of the b state.

\begin{figure}[tb]
\includegraphics[width=7.6cm]{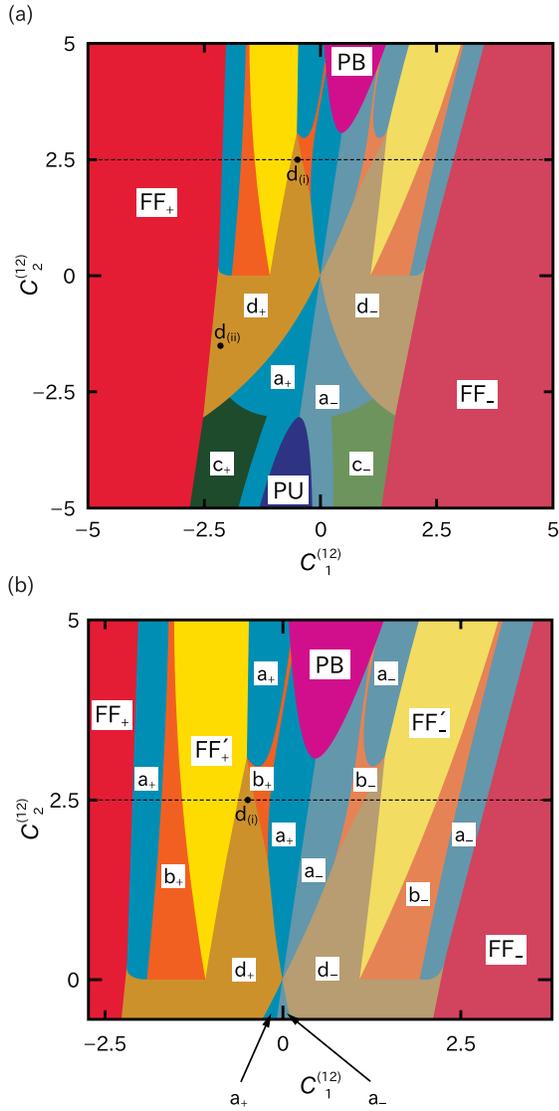}
\caption{
  (color online) Ground-state phase diagram for $c_1^{(1)} = -0.46$,
  $c_1^{(2)} = 1.1$, and $c_2^{(2)} = 1.5$.
  The ground state for $c_1^{(12)} = c_2^{(12)} = 0$ is the ferromagnetic
  state for spin 1 and the cyclic state for spin 2.
  The region of many phases in (a) is magnified in (b).
  The physical quantities along the dotted line are shown in
  Fig.~\ref{f:ferro_cy2}(a).
  The spin states at the black dots are shown in
  Fig.~\ref{f:ferro_cy2}(b).
 }
\label{f:ferro_cy}
\end{figure}
\begin{figure}[tb]
\includegraphics[width=8.5cm]{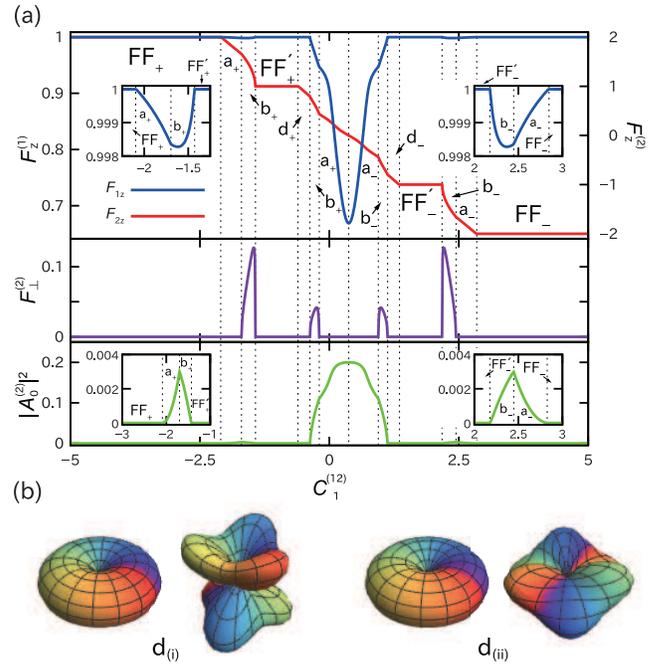}
\caption{
  (color online) (a) Dependence of $F_z^{(1)}$, $F_z^{(2)}$,
  $F_\perp^{(2)}$, and $|A_0^{(2)}|^2$ on $c_1^{(12)}$ along the dotted line
  in Fig.~\ref{f:ferro_cy}.
  Here, the spin-1 and spin-2 states are rotated so that $\bm{F}^{(1)}$ is
  in the $z$ direction, and hence $F_\perp^{(1)}$ is always zero.
  The small changes in $F_z^{(1)}$ and $|A_0^{(2)}|$ are magnified in the
  insets.
  (b) Spherical-harmonic representations of the spin states marked by the
  black dots in Fig.~\ref{f:ferro_cy}, where the left- and right-hand
  figures are the spin-1 and spin-2 states, respectively.
 }
\label{f:ferro_cy2}
\end{figure}
Figure~\ref{f:ferro_cy} shows the ground-state phase diagram for
$c_1^{(1)} = -0.46$, $c_1^{(2)} = 1.1$, and $c_2^{(2)} = 1.5$.
If the inter-spin interaction is absent, i.e., at the origin of the phase
diagram, the ground state of the spin-1 BEC is the ferromagnetic state and
that of the spin-2 BEC is the cyclic state for these parameters.
The phase diagram is again very complicated.
Let us examine the phases along the dotted line.
As $c_1^{(12)}$ is increased from a large negative value, the ground state
changes from the FF$_+$ state to the a$_+$, b$_+$, and FF$'_+$ states, which
is similar to the case in Fig.~\ref{f:ferro_nem}.
After that, a new phase appears, labeled by d$_+$.
In this phase, the value of $|A_0^{(2)}|$ in the spin-2 state vanishes, as
in the cyclic state, whereas $|\bm{F}^{(2)}|$ is finite, as shown in
Fig.~\ref{f:ferro_cy2}(a).
The spin-1 state is in the ferromagnetic state $\bm{\zeta}^{(1)} =
\bm{\zeta}^{(1)}_F$.
From the shape of the spherical harmonic representation in
Fig.~\ref{f:ferro_cy2}(b), we find that this state may be regarded as an
intermediate state between the FC and FF$'$ states.
The d$_\pm$ states also exist in the region $c_2^{(12)} < 0$.
The structures of the a$_\pm$, b$_\pm$, and c$_\pm$ regions in
Fig.~\ref{f:ferro_cy} appear to be different from those in
Fig.~\ref{f:ferro_nem}.

\begin{figure}[tb]
\includegraphics[width=8cm]{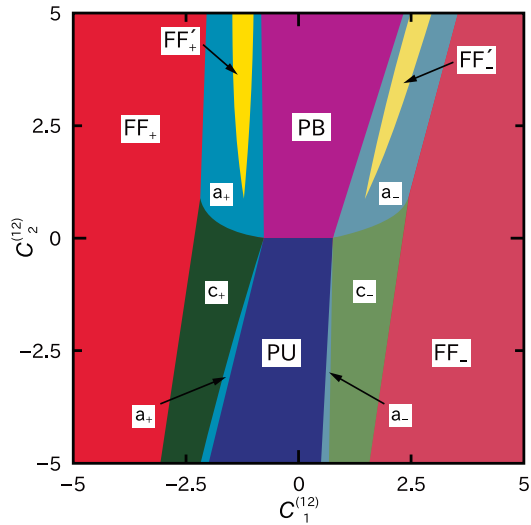}
\caption{
  (color online) Ground-state phase diagram for $c_1^{(1)} = 0.46$,
  $c_1^{(2)} = 1.1$, and $c_2^{(2)} = -1$.
  The ground state for $c_1^{(12)} = c_2^{(12)} = 0$ is the polar state for
  spin 1 and the nematic state for spin 2.
 }
\label{f:pol_nem}
\end{figure}
Figure~\ref{f:pol_nem} shows the ground-state phase diagram for
$c_1^{(1)} = 0.46$, $c_1^{(2)} = 1.1$, and $c_2^{(2)} = -1$.
If the inter-spin interaction is absent, the ground state of the spin-1 BEC
is the polar state and that of the spin-2 BEC is the nematic state for these
parameters.
We find from Fig.~\ref{f:pol_nem} that the PB and PU phases extend and
contact each other at $c_2^{(12)} = 0$.
In this phase diagram there is no symmetry broken state, such as the b
state.

\begin{figure}[tb]
\includegraphics[width=7.6cm]{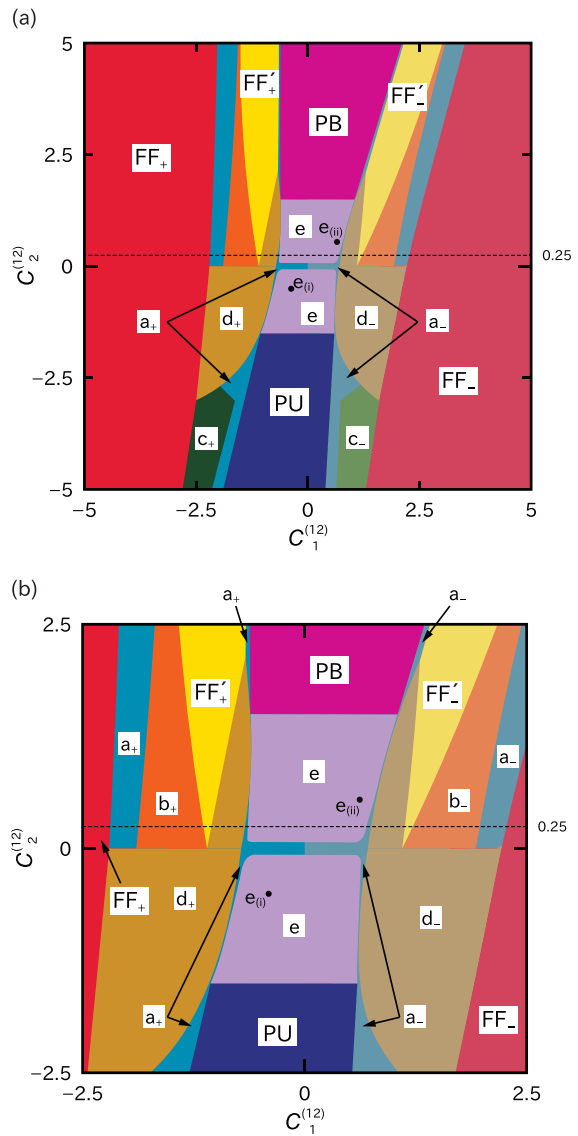}
\caption{
  (color online) Ground-state phase diagram for $c_1^{(1)} = 0.46$,
  $c_1^{(2)} = 1.1$, and $c_2^{(2)} = 1.5$.
  The ground state for $c_1^{(12)} = c_2^{(12)} = 0$ is the polar state for
  spin 1 and the cyclic state for spin 2.
  The region of many phases in (a) is magnified in (b).
  The physical quantities along the dotted line in (a) are shown in
  Fig.~\ref{f:pol_cy2}(a).
  The spin states at the black dots are shown in
  Fig.~\ref{f:pol_cy2}(b).
 }
\label{f:pol_cy}
\end{figure}
\begin{figure}[tb]
\includegraphics[width=8.5cm]{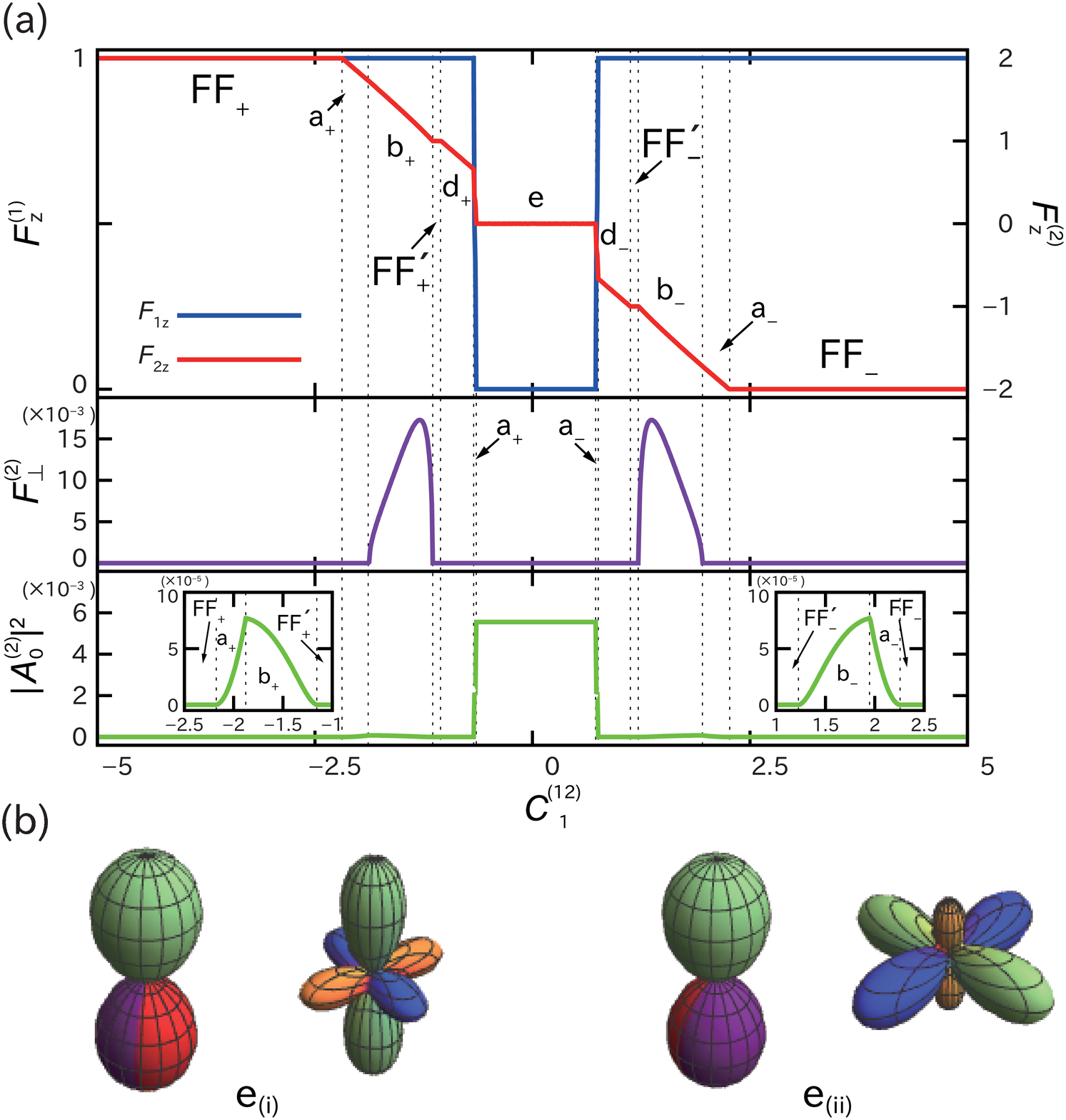}
\caption{
  (color online) (a) Dependence of $F_z^{(1)}$, $F_z^{(2)}$,
  $F_\perp^{(2)}$, and $|A_0^{(2)}|^2$ on $c_1^{(12)}$ along the dotted line
  in Fig.~\ref{f:pol_cy}.
  Here, the spin-1 and spin-2 states are rotated so that $\bm{F}^{(1)}$ is
  in the $z$ direction, and hence $F_\perp^{(1)}$ is always zero.
  The small changes in $F_z^{(1)}$ and $|A_0^{(2)}|$ are magnified in the
  insets.
  (b) Spherical-harmonic representations of the spin states marked by the
  black dots in Fig.~\ref{f:ferro_cy}, where the left- and right-hand
  figures are the spin-1 and spin-2 states, respectively.
 }
\label{f:pol_cy2}
\end{figure}
Figure~\ref{f:pol_cy} shows the ground-state phase diagram for $c_1^{(1)} =
0.46$, $c_1^{(2)} = 1.1$, and $c_2^{(2)} = 1.5$.
If the inter-spin interaction is absent, the ground state of the spin-1 BEC
is the polar state and that of the spin-2 BEC is the cyclic state for these
parameters.
In this phase diagram, a new state appears, labeled e.
The e state has no magnetization for both spin 1 and spin 2, $\bm{F}^{(1)} =
\bm{F}^{(2)} = 0$, as shown in Fig.~\ref{f:pol_cy2}(a).
From the shapes of the spherical harmonic representation in
Fig.~\ref{f:pol_cy2}(b), the e state is an intermediate state between the
cyclic and nematic states.
In the phase diagram, the regions of the e state are located at the heads of 
the PB and PU regions.
For the parameters in Fig.~\ref{f:pol_cy}, interestingly, the two regions of
the e state are detached from each other near the origin, where the a$_\pm$
states fill in.
Although in Fig.~\ref{f:pol_cy2} the quantities $\bm{F}^{(1)}$,
$\bm{F}^{(2)}$, and $A_0^{(2)}$ seem to jump at the boundary of the e
region, they continuously change across the very narrow regions of the
a$_\pm$ states.
In all of the phase diagrams presented above, these quantities continuously
change at the phase boundaries of the intermediate (a, b, c, d, and e)
regions.

\begin{figure}[tb]
\includegraphics[width=8cm]{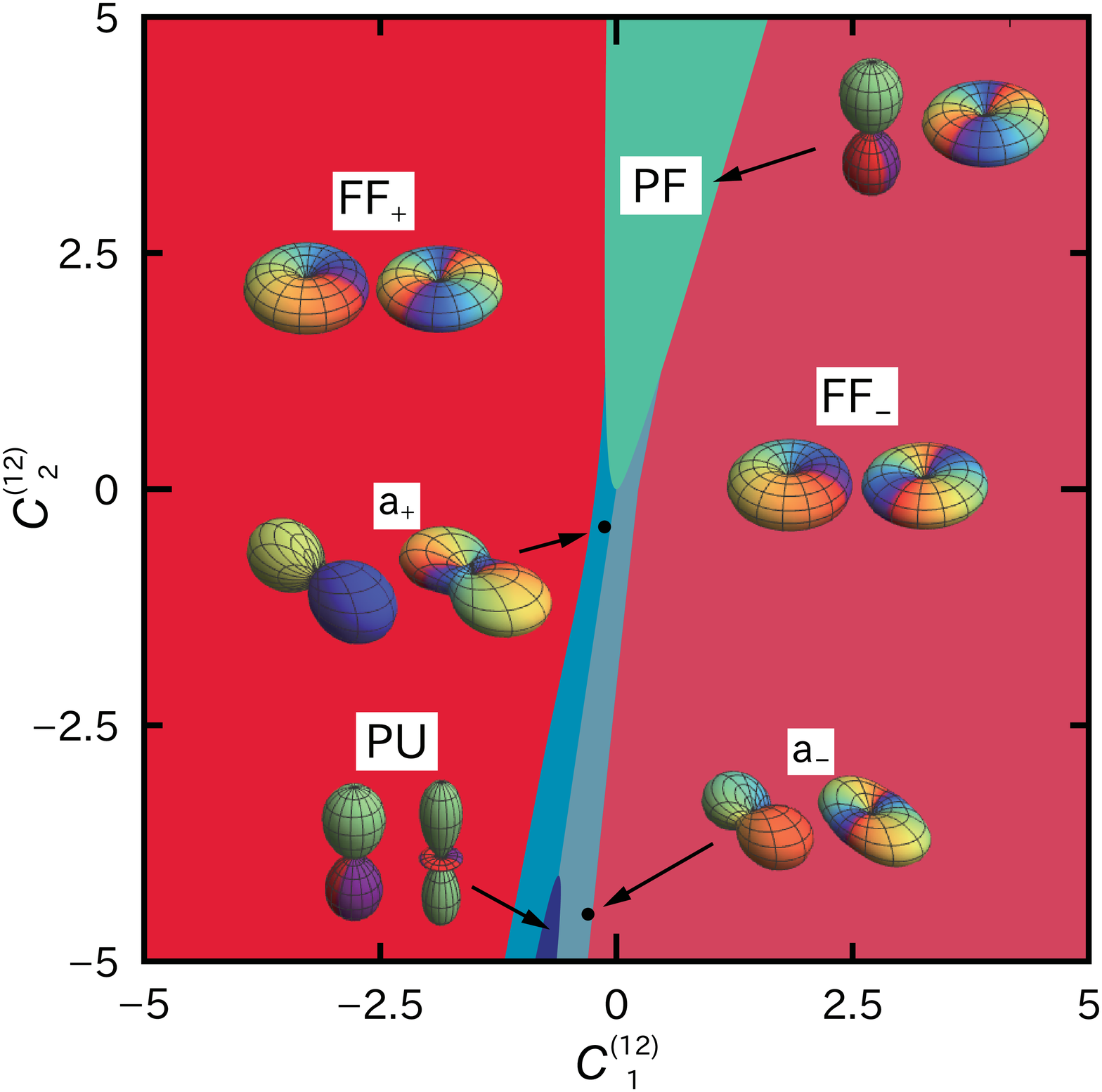}
\caption{
  (color online) Ground-state phase diagram for $c_1^{(1)} = 0.46$,
  $c_1^{(2)} = -0.0005$, and $c_2^{(2)} = 1$.
  The ground state for $c_1^{(12)} = c_2^{(12)} = 0$ is the polar state for
  spin 1 and the ferromagnetic state for spin 2.
  The spherical harmonic representations of the spin states are also shown,
  where the left- and right-hand figures represent the spin-1 and spin-2
  states, respectively.
 }
\label{f:pol_ferro}
\end{figure}
Figure~\ref{f:pol_ferro} shows the ground-state phase diagram for $c_1^{(1)}
= 0.46$, $c_1^{(2)} = -0.0005$, and $c_2^{(2)} = 1$.
If the inter-spin interaction is absent, the ground state of the spin-1 BEC
is the polar state and that of the spin-2 BEC is the ferromagnetic state for
these parameters.
We take the small value of $c_1^{(2)}$, because the PU region is far from
the origin for a larger value of $c_1^{(2)}$.
The a$_\pm$ states occupy the region near the origin instead of the PU
state.
Compared with Fig.~\ref{f:noint}, the degeneracy is removed and the PF state
remains in the upper region of Fig.~\ref{f:pol_ferro}.

Finally, we mention the order-parameter manifold of the ground-state.
In the case of individual spin-1 and spin-2 BECs, the Hamiltonian is
invariant with respect to changes in the global phase, U(1), and the
rotation in the spin space, SO(3).
The ground state therefore has continuous degeneracy, with a manifold
represented by U(1) $\times$ SO(3).
However, for example, the spin-1 ferromagnetic state in
Fig.~\ref{f:spin12}(a) is invariant with respect to rotation around the
symmetry axis (with a global phase shift due to the spin-gauge symmetry).
In other words, the isotropy group of the spin-1 ferromagnetic state is
SO(2).
The order-parameter manifold of the spin-1 ferromagnetic state is thus U(1)
$\times$ SO(3) / SO(2) $\simeq$ SO(3)~\cite{Ho}.
The isotropy group of the spin-1 polar state is SO(2) $\times
\mathbb{Z}_2$, since Fig.~\ref{f:spin12}(b) is invariant with respect to
rotation around the symmetry axis and upside-down rotation with global
phase $\pi$.

In the case of the spin-1/spin-2 mixture, the Hamiltonian is invariant with
respect to changes in the global phase for each of the spin-1 and spin-2
states, in addition to the spin rotation of both spin-1 and spin-2 states,
and then the symmetry group of the Hamiltonian is U(1) $\times$ U(1)
$\times$ SO(3).
For example, the isotropy group of the FF state is SO(2), and therefore the
order-parameter manifold of the FF state is U(1) $\times$ SO(3).
Similarly, the FF$'$ and FU states have this manifold.
The isotropy groups of the intermediate states are summarized in
Table~\ref{t:abc}, whose symmetries are lower than those of individual spin
states.
For example, the symmetry-broken state b in Table~\ref{t:abc} only has the
trivial isotropy group (only the identity element).

\section{Conclusions}
\label{s:conc}

We have investigated the ground-state phase diagrams of a mixture of spin-1
and spin-2 BECs in the mean-field approximation.
We obtained two types of ground states.
One is a pair of known stationary states in spin-1 and spin-2 BECs,
such as the FF and PB states.
In the other type of ground state, either or both of the spin states
continuously change with respect to the interaction coefficients.
The latter type of ground state is classified in Table~\ref{t:abc}.

For the various choices of the intra-spin interaction coefficients,
$c_1^{(1)}$, $c_1^{(2)}$, and $c_2^{(2)}$, we obtained the phase diagrams
with respect to the inter-spin interaction coefficients, $c_1^{(12)}$ and
$c_2^{(12)}$.
These phase diagrams have remarkably rich structures.
In all the phase diagrams, the FF$_+$ and FF$_-$ phases occupy the regions
of large negative and positive $c_1^{(12)}$, respectively.
Also, the PF, or the PB and FF$'_\pm$ phases are located in the $c_2^{(12)}
> 0$ region, and the PU phase is located in the $c_2^{(12)} < 0$ region
(except Fig.~\ref{f:ferro_ferro}).
Between these phases, there exist various intermediate phases with
interesting phase structures.
Among them, we found the axisymmetry broken phase (b in Table.~\ref{t:abc}),
in which the spin-1 and spin-2 vectors are tilted from each other.

We have also determined the ground-state phase of a mixture of spin-1 and
spin-2 $^{87}$Rb BECs, using the measured interaction
coefficients~\cite{Eto}.
It has been known that the ground state of the spin-1 $^{87}$Rb BEC alone is
the ferromagnetic state and that of spin-2 BEC is a linear combination of
the uniaxial and biaxial nematic states at zero magnetic field.
By contrast, for an almost 1:1 mixture, the ground state is the polar
state for spin 1 and the biaxial-nematic state for spin 2.
The ground state of the spinor mixture of $^{87}$Rb BECs is thus changed by
the interaction between spin-1 and spin-2 BECs.

The present study can be extended in various directions.
For example, the magnetic field dependence (linear and quadratic) of the
phase diagrams is the next planned extension of this work.
Since the ground-state manifolds of the spinor mixture are different from
those of single BECs, novel topological excitations will be possible.
If phase separation occurs in the spinor mixture, we expect that the
interface between domains will create interesting problems.

\begin{acknowledgments}
This work was supported by JSPS KAKENHI Grant Numbers JP17K05595,
JP17K05596, JP25103007, JP16K05505, and JP15K05233.
YE acknowledges support by Leading Initiative for Excellent Young
Researchers (LEADER).
\end{acknowledgments}

\appendix

\section{Derivation of interaction energy between spin-1 and spin-2 atoms}

The spin state of colliding spin-1 and spin-2 atoms can be represented by
the bases as
\begin{equation}
| {\cal F, M} \rangle = \sum_{mm'} C_{mm'}^{\cal FM}
|1, m \rangle |2, m' \rangle,
\end{equation}
where $C_{mm'}^{\cal FM}$ is the Clebsch-Gordan coefficient, ${\cal F} =
1$, 2, and 3 are total spin, and ${\cal M} = -{\cal F}, -{\cal F} + 1,
\cdots, {\cal F}$.
The projection operator for the colliding channel of total spin ${\cal F}$
is defined by
\begin{equation}
\hat P_{\cal F} = \sum_{{\cal M} = -{\cal F}}^{\cal F}
| {\cal F, M} \rangle \langle {\cal F, M} |,
\end{equation}
which is rotation invariant.
In the present Hilbert space, the identity operator $\hat I$ is given by
\begin{equation} \label{iden}
\hat P_1 + \hat P_2 + \hat P_3 = \hat I.
\end{equation}
We define the spin operators acting on the spin-1 and spin-2 states as
$\hat{\bm{f}}_1$ and $\hat{\bm{f}}_2$, respectively.
We find
\begin{eqnarray} \label{ff}
\hat{\bm{f}}_1 \cdot \hat{\bm{f}}_2 & = &
\frac{1}{2} \left( \hat{\bm{f}}_1 + \hat{\bm{f}}_2 \right)^2
- \frac{1}{2} \sum_{f = 1, 2} f (f + 1) \hat I
\nonumber \\
& = & \frac{1}{2} \sum_{{\cal F} = 1, 2, 3} {\cal F} ({\cal F} + 1)
\hat P_{\cal F} - 4 \hat I.
\end{eqnarray}
Since the Hamiltonian must be rotation invariant, the two-body interaction
Hamiltonian between spin-1 and spin-2 atoms is written as
\begin{equation}
\hat H_{12} = \frac{2\pi\hbar^2}{M_{12}} \sum_{{\cal F} = 1, 2, 3}
a_{\cal F} \hat P_{\cal F} \delta(\bm{r}_1 - \bm{r}_2),
\end{equation}
where $M_{12} = (M_1^{-1} + M_2^{-1})^{-1}$ is the reduced mass.
Using Eqs.~(\ref{iden}) and (\ref{ff}), the interaction Hamiltonian can be
rewritten as
\begin{equation}
\hat H_{12} = \left( g_0^{(12)} \hat I + g_1^{(12)}
\hat{\bm{f}}_1 \cdot \hat{\bm{f}}_2 + g_2^{(12)} \hat P_1 \right)
\delta(\bm{r}_1 - \bm{r}_2),
\end{equation}
where $g_0^{(12)}$, $g_1^{(12)}$, and $g_2^{(12)}$ are defined in
Eq.~(\ref{g12}).
The mean-field energy is thus given by Eq.~(\ref{E12}), where $P_1^{(12)}
= |A_{1,1}|^2 + |A_{1,0}|^2 + |A_{1,-1}|^2$ with
\begin{subequations} \label{P1}
\begin{eqnarray}
A_{1,1} & = & \frac{1}{\sqrt{10}} \zeta_1^{(1)} \zeta_0^{(2)}
- \sqrt{\frac{3}{10}} \zeta_0^{(1)} \zeta_1^{(2)}
+ \sqrt{\frac{3}{5}} \zeta_{-1}^{(1)} \zeta_2^{(2)},
\nonumber \\ \\
A_{1,0} & = & \sqrt{\frac{3}{10}} \zeta_1^{(1)} \zeta_{-1}^{(2)}
- \sqrt{\frac{2}{5}} \zeta_0^{(1)} \zeta_0^{(2)}
+ \sqrt{\frac{3}{10}} \zeta_{-1}^{(1)} \zeta_1^{(2)},
\nonumber \\ \\
A_{1,-1} & = & \sqrt{\frac{3}{5}} \zeta_1^{(1)} \zeta_{-2}^{(2)}
- \sqrt{\frac{3}{10}} \zeta_0^{(1)} \zeta_{-1}^{(2)}
+ \frac{1}{\sqrt{10}} \zeta_{-1}^{(1)} \zeta_0^{(2)}.
\nonumber \\
\end{eqnarray}
\end{subequations}

\section{Linear stability analysis and phase boundaries}

We perform a linear stability analysis of a stationary state to obtain the
phase boundaries analytically.
The total energy is given by
\begin{eqnarray} \label{appE}
E & = & \frac{c_0^{(1)}}{2} \left( \sum_{m=-1}^1 |\zeta_m^{(1)}|^2 \right)^2
+ \frac{c_0^{(2)}}{2} \left( \sum_{m=-2}^2 |\zeta_m^{(2)}|^2 \right)^2
\nonumber \\
& & + \frac{1}{2} \left(
c_1^{(1)} \bm{F}^{(1)} \cdot \bm{F}^{(1)} + c_1^{(2)} \bm{F}^{(2)} \cdot
\bm{F}^{(2)} + c_2^{(2)} |A_0^{(2)}|^2 \right)
\nonumber \\
& & + c_1^{(12)} \bm{F}^{(1)} \cdot \bm{F}^{(2)} + c_2^{(12)} P_1^{(12)}.
\end{eqnarray}
Using this energy, the Gross-Pitaevskii (GP) equation is written as
\begin{equation} \label{GPapp}
i \hbar \frac{\partial \zeta_m^{(f)}}{\partial t} = \frac{\partial
E}{\partial \zeta_m^{(f)*}}.
\end{equation}
All of the ground states in the phase diagrams are stationary solutions
of the GP equation.
We write a stationary solution as
\begin{equation}
\zeta_m^{(f)}(t) = e^{-i \mu_f t / \hbar} Z_m^{(f)},
\end{equation}
where $\mu_f$ is the chemical potential for spin $f$.
We consider a small deviation from the stationary solution as
\begin{equation}
  \zeta_m^{(f)}(t) = e^{-i \mu_f t / \hbar} \left( Z_m^{(f)} + u_m^{(f)}
  e^{-i \omega t} + v_m^{(f)*} e^{i \omega^* t} \right).
\end{equation}
Substituting this into Eq.~(\ref{GPapp}) and taking the first-order terms
of $u_m^{(f)}$ and $v_m^{(f)}$, we obtain an $8 \times 8$ eigenvalue
equation with respect to $\omega$.
If one or more eigenvalues are negative or complex, the stationary state
$Z_m^{(f)}$ is not the ground state.

For example, we take the stationary state $Z_m^{(f)}$ as the ferromagnetic
state $\bm{\zeta}^{(1)} = (1, 0, 0)$ and $\bm{\zeta}^{(2)} = (1, 0, 0, 0,
0)$, which corresponds to the FF$_+$ state in the phase diagrams.
Diagonalizing the eigenvalue equation, we obtain
\begin{subequations}
\begin{eqnarray}
\label{ff2}
\omega & = & -3 c_1^{(12)}, \\
\label{ff3}
\omega & = & -6 c_1^{(2)} - 3 c_1^{(12)} + \frac{3}{10} c_2^{(12)}, \\
\label{ff4}
\omega & = & -8 c_1^{(2)} + \frac{2}{5} c_2^{(2)} - 4 c_1^{(12)} +
\frac{3}{5} c_2^{(12)}, \\
\label{ff5}
  \omega & = & -c_1^{(1)} - 2 c_1^{(2)} - 3 c_1^{(12)} + \frac{7}{20}
  c_2^{(12)} \nonumber \\
  & & \pm \left[ A^2 - \frac{1}{2} A c_2^{(12)} + \left( \frac{7}{20}
    c_2^{(12)} \right)^2 \right]^{1/2},
\end{eqnarray}
\end{subequations}
and $\omega = 0$, where $A = c_1^{(1)} - 2 c_1^{(2)} + c_1^{(12)}$.
In the case of Fig.~\ref{f:noint}, for example, the condition $\omega > 0$
for Eqs.~(\ref{ff2}) and (\ref{ff4}) gives $c_1^{(12)} < 0$ and $c_2^{(12)}
> 10 c_2^{(12)}$, which agree with the phase boundary of the FF$_+$ phase
in Fig.~\ref{f:noint}.
On the other hand, for the phase diagram in Fig.~\ref{f:ferro_nem}, the
phase boundary of the FF$_+$ phase is determined by Eqs.~(\ref{ff4}) and
(\ref{ff5}) for $c_2^{(12)} < 0$ and $c_2^{(12)} > 0$, respectively.

Taking the stationary state $Z_m^{(f)}$ as $\bm{\zeta}^{(1)} = (1, 0, 0)$
and $\bm{\zeta}^{(2)} = (0, 0, 0, 0, 1)$, i.e., the FF$_-$ state, we obtain
\begin{subequations}
\begin{eqnarray}
\label{ffm1}
\omega & = & -6 c_1^{(2)} + 3 c_1^{(12)} - \frac{3}{5} c_2^{(12)}, \\
\label{ffm2}
\omega & = & -8 c_1^{(2)} + \frac{2}{5} c_2^{(2)} + 4 c_1^{(12)} -
\frac{3}{5} c_2^{(12)}, \\
\label{ffm5}
\omega & = & \pm \left( -c_1^{(1)} + 2 c_1^{(2)} + c_1^{(12)} -
\frac{1}{20} c_2^{(12)} \right) \nonumber \\
& & + \left[ B^2 - \frac{11}{10} B c_2^{(12)} + \frac{97}{400} c_2^{(12)2}
\right]^{1/2}, \\
\label{fm7}
\omega & = & \left| c_1^{(12)} - \frac{3}{10} c_2^{(12)} \right|,
\end{eqnarray}
\end{subequations}
and $\omega = 0$, where $B = c_1^{(1)} + 2 c_1^{(2)} - 3 c_1^{(12)}$.
For example, for the phase diagram in Fig.~\ref{f:ferro_nem}, the phase
boundary of the FF$_-$ phase is determined by Eqs.~(\ref{ffm2}) and
(\ref{ffm5}) for $c_2^{(12)} < 0$ and $c_2^{(12)} > 0$, respectively.

\end{document}